\documentclass[%
aps,
prb,
 amsmath,amssymb,
 superscriptaddress,
 reprint,%
]{revtex4-1}

\usepackage{graphicx}
\usepackage{dcolumn}
\usepackage{bm}
\usepackage{ulem}
\usepackage[utf8]{inputenc}
\usepackage[T1]{fontenc}
\usepackage{mathptmx}
\usepackage{textcomp} 
\usepackage{nicefrac}
\usepackage{xcolor}
\usepackage{gensymb}

\begin{document}

\preprint{AIP/123-QED}

\title[Direct Access to Auger recombination in Graphene]{Direct Access to Auger Recombination in Graphene}


\author{Marius Keunecke} \email{mkeunec@gwdg.de}%
\address{I. Physikalisches Institut, Georg-August-Universit\"at G\"ottingen, Friedrich-Hund-Platz 1, 37077 G\"ottingen, Germany}

\author{David Schmitt} %
\address{I. Physikalisches Institut, Georg-August-Universit\"at G\"ottingen, Friedrich-Hund-Platz 1, 37077 G\"ottingen, Germany}

\author{Marcel Reutzel} \email{marcel.reutzel@phys.uni-goettingen.de}%
\address{I. Physikalisches Institut, Georg-August-Universit\"at G\"ottingen, Friedrich-Hund-Platz 1, 37077 G\"ottingen, Germany}

\author{Marius Weber} %
\address{TU Kaiserslautern, Erwin-Schrödinger-Str. 46, 67663 Kaiserslautern, Germany}

\author{Christina Möller}%
\address{I. Physikalisches Institut, Georg-August-Universit\"at G\"ottingen, Friedrich-Hund-Platz 1, 37077 G\"ottingen, Germany}

\author{G.~S.~Matthijs~Jansen} %
\address{I. Physikalisches Institut, Georg-August-Universit\"at G\"ottingen, Friedrich-Hund-Platz 1, 37077 G\"ottingen, Germany}

\author{Tridev~A.~Mishra}%
\affiliation{Institut für Theoretische Physik, Georg-August-Universit\"at G\"ottingen, Friedrich-Hund-Platz 1, 37077 G\"ottingen, Germany}

\author{Alexander~Osterkorn}%
\affiliation{Institut für Theoretische Physik, Georg-August-Universit\"at G\"ottingen, Friedrich-Hund-Platz 1, 37077 G\"ottingen, Germany}

\author{Wiebke~Bennecke}%
\affiliation{I. Physikalisches Institut, Georg-August-Universit\"at G\"ottingen, Friedrich-Hund-Platz 1, 37077 G\"ottingen, Germany}

\author{Klaus Pierz} %
\address{Physikalisch-Technische Bundesanstalt, Bundesallee 100, 38116 Braunschweig, Germany}

\author{Hans Werner Schumacher} %
\address{Physikalisch-Technische Bundesanstalt, Bundesallee 100, 38116 Braunschweig, Germany}

\author{Davood Momeni Pakdehi} %
\address{Physikalisch-Technische Bundesanstalt, Bundesallee 100, 38116 Braunschweig, Germany}


\author{Daniel Steil} %
\address{I. Physikalisches Institut, Georg-August-Universit\"at G\"ottingen, Friedrich-Hund-Platz 1, 37077 G\"ottingen, Germany}

\author{Salvatore~R.~Manmana}%
\affiliation{Institut für Theoretische Physik, Georg-August-Universit\"at G\"ottingen, Friedrich-Hund-Platz 1, 37077 G\"ottingen, Germany}

\author{Sabine Steil} 
\address{I. Physikalisches Institut, Georg-August-Universit\"at G\"ottingen, Friedrich-Hund-Platz 1, 37077 G\"ottingen, Germany}

\author{Stefan~Kehrein}%
\affiliation{Institut für Theoretische Physik, Georg-August-Universit\"at G\"ottingen, Friedrich-Hund-Platz 1, 37077 G\"ottingen, Germany}

\author{Hans Christian Schneider} %
\address{TU Kaiserslautern, Erwin-Schrödinger-Str. 46, 67663 Kaiserslautern, Germany}

\author{Stefan Mathias} \email{smathias@uni-goettingen.de}%
\address{I. Physikalisches Institut, Georg-August-Universit\"at G\"ottingen, Friedrich-Hund-Platz 1, 37077 G\"ottingen, Germany}

\begin{abstract}
Auger scattering channels are of fundamental importance to describe and understand the non-equilibrium charge carrier dynamics in graphene. While impact excitation increases the number of carriers in the conduction band and has been observed experimentally, direct access to its inverse process, Auger recombination, has so far been elusive.
Here, we tackle this problem 
by applying our novel setup for ultrafast time-resolved photoelectron momentum microscopy. Our approach gives simultaneous access to charge carrier dynamics at all energies and in-plane momenta within the linearly dispersive Dirac cones. We thus provide direct evidence for Auger recombination on a sub-10~fs timescale by identifying transient energy- and momentum-dependent populations far above the excitation energy. We compare our results with model calculations of scattering processes in the Dirac cone to support our experimental findings.


\end{abstract}

\maketitle


Non-equilibrium light-matter interaction processes have been studied in graphene as a prototype system for fundamental energy dissipation channels of non-thermal and hot charge carriers in two-dimensional systems~\cite{Malic17adp, rana2007electron, dawlaty2008measurement, Malic11prb, Winnerl11prl, Breusing11prb, strait2011very, trushin2011anisotropic,Winzer12prb, song2013photoexcited, brida2013ultrafast, Gierz13natmat, Johannsen13prl, svintsov2014carrier, mittendorff2014anisotropy, kadi2015impact, Gierz15prl, Johannsen15nanolett, konig2016slow, Winzer16prb, aeschlimann_ultrafast_2017, Tan17prx, Rode18prl, Na19sci, Caruso20prb}. In order to access these out-of-equilibrium properties on the femtosecond timescale, the optically-excited non-thermal charge carrier distributions are commonly probed using ultrafast optical~\cite{Malic11prb, Winnerl11prl, Breusing11prb, brida2013ultrafast, konig2016slow, strait2011very, dawlaty2008measurement} and photoemission~\cite{Gierz13natmat, Johannsen13prl, Gierz15prl, Johannsen15nanolett, aeschlimann_ultrafast_2017, Tan17prx, Rode18prl, Na19sci} spectroscopy. Due to the linear band dispersions with a vanishing density of states at the Dirac point combined with the weak screening of the Coulomb interaction in the two-dimensional material, many-particle electron-electron (e-e) interactions are particularly strong. As a result, the excited charge carriers thermalize to a hot Fermi-Dirac distribution on the $\approx$50~fs timescale~\cite{Rode18prl, Johannsen15nanolett, brida2013ultrafast}. Concurrently, scattering with optical phonons leads to an azimuthal thermalization and further cooling of the Fermi-Dirac distribution on a timescale of 200~femtoseconds to a few picoseconds~\cite{mittendorff2014anisotropy, Johannsen13prl,Gierz13natmat, Johannsen15nanolett}. Of particular interest in the carrier thermalization and cooling dynamics in graphene are Auger processes, where one carrier changes from the valence to the conduction band and vice versa, while the other involved carrier remains in the same band. Here, one distinguishes impact excitation (IE) and its inverse process Auger recombination (AR), which increase and decrease the number of carriers in the conduction band, respectively. For example, Auger processes can lead to a population inversion in the conduction and the valence band, highlighting its potential as a gain medium in optoelectronic devices~\cite{Sun10acsnano, winnerl2017ultrafast}. However, while IE has been clearly identified experimentally in graphene~\cite{Winzer12prb, Gierz15prl} and other materials~\cite{Mathias:2016ey}, to the best of our knowledge, a direct experimental verification and analysis of AR processes have remained elusive.


\begin{figure}[hbt!]
    \centering
    \includegraphics[width=1\linewidth]{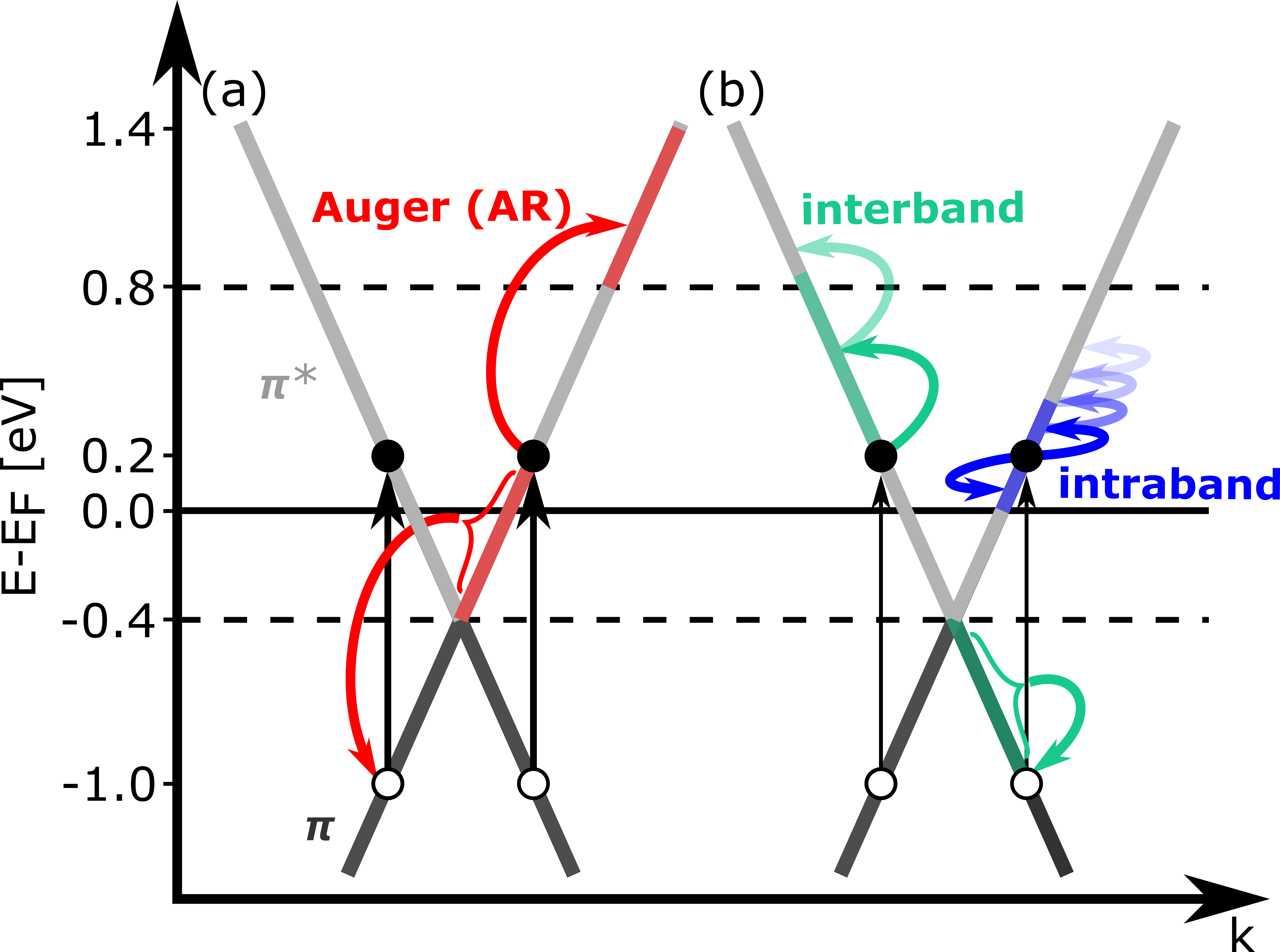}
    \caption{Possible two-body e-e scattering events of non-thermal charge carriers in n-doped graphene that are able to create population at energies higher than reached by the excitation itself. Excitation with 1.2~eV pump photons (black arrows) resonantly populates states around $E-E_F=0.2$~eV (black dots). Two processes are distinguished: (a) Band-crossing scattering processes via Auger recombination. Here, one electron bridges the valence and the conduction band and the other involved electron gains the energy of the recombination process and can reach energies up to about 1.2~eV above the Fermi level. (b) Multiple inter- and intraband scattering processes (green and blue arrows, respectively), which can occupy higher energies in the conduction band via cascaded scattering processes.}
        \label{fig:scattering_intro}
\end{figure}

In this manuscript, we provide direct experimental evidence for AR on timescales as short as 10~fs. We show that AR induces an energy- and momentum-dependent population in the conduction band at energies higher than that reached by the optical excitation itself. Moreover, depending on the efficiency of AR vs. other scattering processes, we find distinct temporal shifts of transient energy- and momentum-dependent carrier populations in the conduction band. Our findings are supported by model calculations of two-body Coulomb scattering processes in graphene.

We specifically selected an n\nobreakdash-doped epitaxial graphene sample~\cite{kruskopf2016comeback, momeni2018minimum} for our study, because it is known that in this case strong optical excitation of charge carriers just above the Fermi-level is most advantageous for dominant Auger recombination processes~\cite{Malic17adp, kadi2015impact, Winzer16prb}. Thus, the Dirac cone is located 0.4~eV below the Fermi-level and excitation with 1.2~eV photons results in a non-equilibrium electron distribution with its peak located at 0.2~eV due to resonant optical excitation ($\approx$0.6~eV above the Dirac point, black arrows in Fig.~\ref{fig:scattering_intro}). During and subsequent to this excitation with a 37$\pm$3~fs laser pulse, multiple two-body scattering processes redistribute the charge carrier population. From these scattering processes, we aim here to identify clear signatures of band-crossing AR, which is generally important for the understanding of non-equilibrium carrier thermalization and cooling in graphene, and, in addition, a process that creates high energy charge carriers during and subsequent to the optical excitation. 

Figure~\ref{fig:scattering_intro} illustrates band-crossing Auger recombination (Fig.~\ref{fig:scattering_intro}a) and inter- and intraband scattering (Fig.~\ref{fig:scattering_intro}b). These particular two-body Coulomb scattering processes can generate population at energies higher than the initial optical excitation, which is the experimental signature we are going to follow, and are therefore relevant for the discussion below. For the Auger recombination process illustrated in Fig.~\ref{fig:scattering_intro}~(a), an electron in the conduction band recombines with a hole in the valence band, and another electron in the conduction band gains the excess energy. In our case, typical energies that can be reached in such a band-crossing Auger scattering event for the scattering partner electron are 1.2~eV above the Fermi level, if this electron originates from energies around the Fermi-level [see red arrows in Fig.~\ref{fig:scattering_intro}~(a)]. Thus, if AR is indeed of importance for the non-equilibrium dynamics in graphene~\cite{Malic17adp, kadi2015impact, Winzer16prb}, we expect such high-energy electrons to be a detectable signature of AR in our time-resolved momentum microscopy experiment. Moreover, the temporal evolution of such a high-energy electron population will give us quantitative information on the Auger recombination scattering rate itself.

We stress that the identification of electrons at energies higher than those reached by the direct optical excitation might not be sufficient proof for the presence of AR. Strong optical excitation, as we carry out in our experiment, apart from creating a pronounced non-equilibrium situation also considerably increases the kinetic energy of the carriers. Coulomb scattering processes will therefore instantly contribute towards establishing a quasi-equilibrium at a higher ``temperature.'' 
This involves cascaded intra- and interband scattering processes that also generate population at higher energies in the conduction band [green and blue arrows in Fig.~\ref{fig:scattering_intro}~(b)]. 
The most important aspect for our experimental study is now to realize that the build-up of high-energy electron populations by multiple cascaded scattering processes exhibits distinct temporal structures that distinguishes it from direct band-crossing Auger recombination processes. In the former case, build-up of population at higher energies requires more and more individual intra- and interband scattering processes: the higher the energy, the longer it takes to create population. In contrast, for AR, energies up to about 1.2~eV can be reached within a single Auger scattering event, and potentially at much earlier times than expected when cascaded multiple intra- and interband scattering events are necessary to reach these energies.


\begin{figure}[hbt!]
    \centering
    \includegraphics[width=\linewidth]{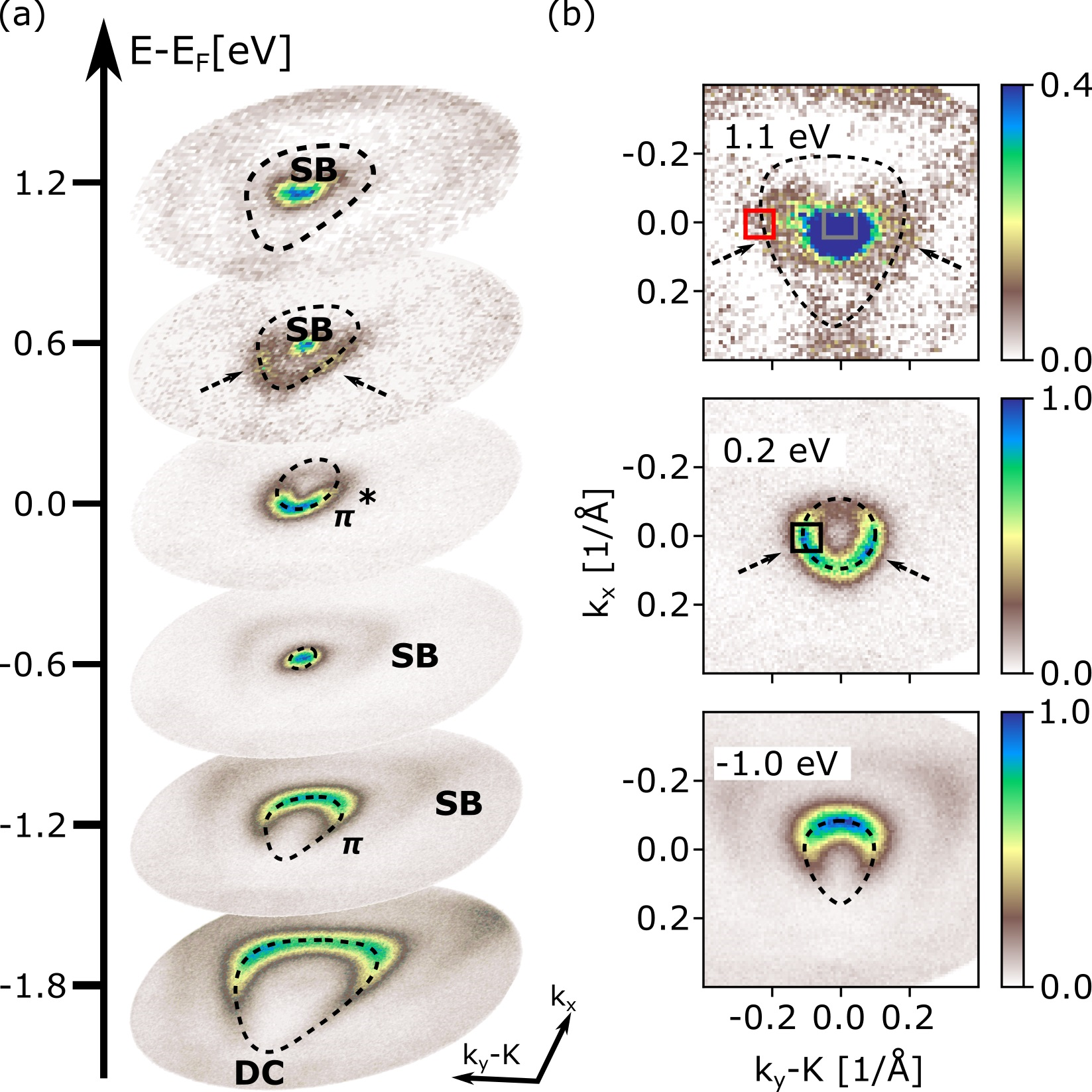}
    \caption{
    Time-of-flight momentum microscopy data obtained on the Dirac cone of n-doped graphene when excited with $p$-polarized IR pump pulses and probed with EUV light ($\Delta t = 10$~fs). (a) 3D illustration of the collected dataset, showing the linear-dispersive valence ($\pi$) and conduction ($\pi^{*}$) bands (marked by dashed lines), the dark corridor (DC), the anisotropic optical excitation (black dashed arrows), and the photon-dressed sideband (SB). The shape of the Dirac cone is indicated by a dotted line. (b) Selected ($k_x$, $k_y$)-momentum maps at $E-E_F=-1.0$~eV, $0.2$~eV, and $1.1$~eV that show cutouts from the multidimensional data set in (a). The momentum-anisotropic charge carrier distribution is conserved up to highest resolved energies $E-E_F=1.1$~eV (black dashed arrows), implicating collinear e-e scattering events to reach these high energies. 
    }
    \label{fig:spectroscopy}
\end{figure}

\begin{figure}[hbt!]
    \centering
    \includegraphics[width=0.5\textwidth]{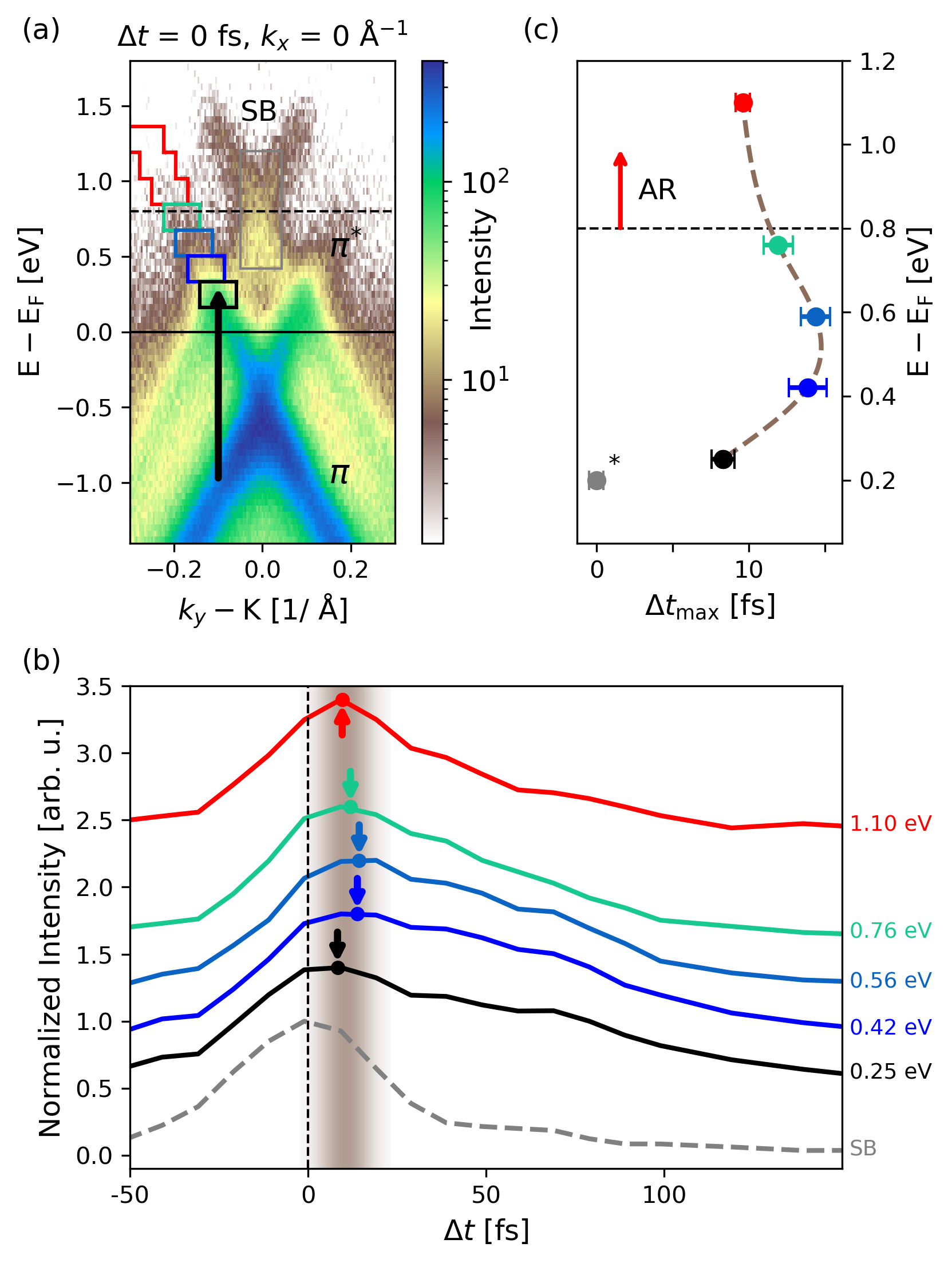}
    \caption{Population dynamics in the conduction band of graphene. (a) ($E$, $k_y$)-cut through the multidimensional data for $k_x=0$. The regions of interest evaluated in (b) and (c) are indicated by colored boxes. The dotted line at $E-E_F=0.8$\,eV indicates the threshold energy that can be reached with a single inter- or intraband scattering event. Higher energies can only be reached by a single AR or multiple inter- and intraband scattering events.
    (b) The photoemission yield in the ROIs as a function of pump-probe delay. The time of maximum charge carrier occupation ($\Delta t_{\rm max}$), obtained by Gaussian fitting of the time traces in the respective ROI between -20\,fs and +30\,fs, is indicated by colored arrows. The grey dashed line corresponds to a cross-correlation of pump an probe pulses, extracted from the time-dependent sideband yield. 
    (c) Energy-dependence of $\Delta t_{\text{max}}$; error bars correspond to the 1$\sigma$ standard deviation of the fit. For increasing $E-E_F$, $\Delta t_{\rm max}$ first increases, saturates, and finally decreases again for energies where AR is expected to become the dominant scattering event. The brown dashed line is a guide to the eye through the data. }  
    \label{fig:dynamics}
\end{figure}

In order to study the temporal structure of charge carrier populations at energies higher than those reached by the direct optical excitation, we now turn to our experimental data collection and analysis. In our time-resolved photoemission experiment, we use a multidimensional data collection scheme that consists of a time-of-flight momentum microscope~\cite{medjanik_direct_2017} in combination with a 1~MHz table-top extreme ultraviolet (EUV) high-harmonic generation beamline (p-polarized, h$\nu$ = 26.5~eV, pulse length $\approx$ 20~fs, angle of incidence = 68\textdegree{})~\cite{Keunecke20timeresolved}. An exemplary time-resolved measurement is shown in  Fig.~\ref{fig:spectroscopy}~(a), which illustrates an ($E$, $k_x$, $k_y$)-resolved data set that is collected at $\Delta t = +10$~fs after the maximum intensity of the pump pulse envelope (p-polarized, h$\nu$ = 1.2~eV, incident fluence: 6.5~mJ/c$\mathrm{m}^2$, pulse length: 37$\pm$3~fs). We have aligned the momentum microscope such that photoelectrons originating from the K-point located in the plane of incidence of the impinging light are imaged onto the center of the photoelectron detector~\cite{Keunecke20timeresolved}. Clearly visible is the ring-like structure of the Dirac cone (black dashed lines as a guide to the eye) below the Fermi-level, and also (at lower intensity) above the Fermi-level due to prior excitation with the 1.2~eV pump pulse [compare also with the ($E$, $k_y$)-map in Fig.~\ref{fig:dynamics}~(a)]. Note that in photoemission some parts of the Dirac cone are not visible due to the so-called "dark corridor of graphene" (marked with "DC")~\cite{Shirley95prb, Gierz11prb}, and that the Dirac cone shows a deviation from the perfect circular shape in the momentum maps, called "trigonal warping"~\cite{ando1998berry, dresselhaus2002intercalation}. In addition, it is known and also seen in our experiment that excitation with 1.2~eV laser pulses is anisotropic (indicated with black dashed arrows for better visibility), which can be explained by a polarization-dependent matrix element for the optical excitation~\cite{trushin2011anisotropic,aeschlimann_ultrafast_2017}. Hence, in our case, the excitation is primarily in $\pm k_y$-direction. All replica features marked with "SB" (for sideband) are induced by the pump pulse, as we discuss in Refs.~\onlinecite{Keunecke20prb, Keunecke20timeresolved}. However, these features (i.e., all features not marked with black dashed lines in Fig.~\ref{fig:spectroscopy}) do not influence or contribute significantly to the observed electron dynamics (see SI).

Turning to the middle panel of Fig.~\ref{fig:spectroscopy}~(b) we find that the highest density of charge carriers in the conduction band above $E_F$ is observed at $E-E_F=0.2$~eV, as expected, because of resonant excitation by 1.2~eV light pulses. As a result, the carrier distribution strongly deviates from a hot Fermi-Dirac distribution, both in its energy and momentum distribution. Most interestingly for our analysis is, however, that we also observe population in the conduction band for energies far above the excitation energy, i.e., up to $E-E_F \approx 1.1$~eV [Fig.~\ref{fig:spectroscopy}~(b), top panel]. Even more, these high-energy charge carriers are also distributed anisotropically in momentum space (indicated again by black dashed arrows). From this observation, we can already draw several conclusions: (i) As these high-energy charge carriers cannot be directly generated by the optical excitation (we exclude nonlinear optical excitation, see SI), we conclude that very efficient scattering events within the timescale of the pump and probe pulses must be responsible for our observation. (ii) The replication of the anisotropic distribution in momentum space is a clear signature that collinear scattering processes must dominate the generation of these high-energy electrons (black dashed arrows in Fig.~\ref{fig:spectroscopy}). (iii) Since these scattering events must be primarily collinear and must have occurred on the few-femtosecond timescale ($\Delta t = +10$~fs in Fig. 2), we can exclude significant contributions of electron-phonon scattering. 


Figure~\ref{fig:dynamics} presents a detailed analysis of the sub-50-fs temporal structure of these high-energy populations, which should be distinct for AR vs.~inter- and intraband scattering processes. In our analysis, we focus on a momentum slice for which $k_x=0$, where the conduction band is directly populated in $\pm k_y$-direction by the optical excitation [compare black dashed arrows in Fig.~\ref{fig:spectroscopy}]. In Fig.~\ref{fig:dynamics}~(a), we show the respective ($E$, $k_y$)-cut and indicate the regions-of-interest (ROI) with colored boxes that are further analyzed in Fig.~\ref{fig:dynamics}~(b). At this point, we make use of $p$-polarized pump light that generates photon-dressed sidebands (SB) and thus provides a direct calibration of the time axis~\cite{saathoff_laser-assisted_2008, Keunecke20prb}: An analysis of the SB intensity yields a direct cross-correlation of the pump and probe laser pulses, implicating that maximum SB intensity is reached when the pump and probe pulses are in temporal overlap [i.e., $\Delta t =0$~fs, grey trace in Fig.~\ref{fig:dynamics}~(b)]. Note that the sideband structure and the main Dirac cone cross in the energy- and momentum-region where the resonant optical excitation occurs [black box in Fig.~\ref{fig:dynamics}~(a)]. However, as we detail in the SI, we can exclude a significant contribution of this sideband yield to the measured dynamics in our further analysis.

In Fig.~\ref{fig:dynamics} we evaluate the occupation of the conduction band as a function of $\Delta t$ separately for each ROI indicated in the ($E$, $k_y$)-resolved data set. 
Interestingly, already for $E-E_F=0.2$~eV (black ROI), where the optical excitation proceeds resonantly, maximum signal is not observed for $\Delta t =0$~fs, but occurs with delay at $\Delta t_{\rm max} \approx 8 \pm 1$~fs. This delayed response of the photoelectron yield can be attributed to finite lifetimes~\cite{Hertel97} due to e-e scattering processes redistributing the charge carrier density already during the pump pulse, which is in agreement with earlier reports~\cite{Winzer16prb}. When evaluating these time traces for increasing energies above the resonant excitation peak, we observe that the maximum population is reached at even larger $\Delta t_{\rm max}$ (blueish traces). At first glance, this is counter-intuitive in terms of excited state lifetimes when considering the relaxation of charge carriers towards the Fermi level, see, e.g. Refs.~\onlinecite{Berthold02prl,Sobota12prl}. However, this behavior should indeed be expected for the generation of this population via cascaded intra- and interband scattering processes: because of energy conservation and the phase space that is available for inter- and intraband scattering events, more and more scattering events are necessary to reach higher and higher energies. Consequently, $\Delta t_{\rm max}$ must increase towards higher energies. However, at the highest energies where we still detect electron population, $\Delta t_{\rm max}$ decreases again (green and red trace). The temporal behavior of the charge carriers at these high energies implies that these carriers have not been excited via multiple scattering events. Instead, they must have been excited to these energies in a single scattering event, which, because of energy conservation, can only be AR. The time to reach maximum population for the highest measured energy can be used for a quantitative estimation of the average Auger recombination time $\tau_{\rm AR}$, and at $E-E_F = 1.1$~eV we extract $\Delta t_{\rm max}$ = $10\pm1$~fs. This value serves as an upper limit, because the observed maximum might additionally be delayed due to finite lifetimes at the respective energies. Therefore, we deduce that the AR time is extremely fast with $\tau_{\rm AR}$ < $10$~fs.
In Ref.~\onlinecite{Gierz15prl}, it was found that the maximum carrier multiplication due to IE was reached within 26~fs of the optical excitation. This requires the average scattering time for IE to be well below this value, placing it also on the 10-fs timescale. Although a direct comparison of our work with Ref.~\onlinecite{Gierz15prl} is hindered by the many parameters which influence Auger scattering times (e.g. phase space, Pauli blocking and screening), we conclude that a sub-10-fs AR time is reasonable.

\begin{figure}[hbt!]
    \centering
    \includegraphics[width=\linewidth]{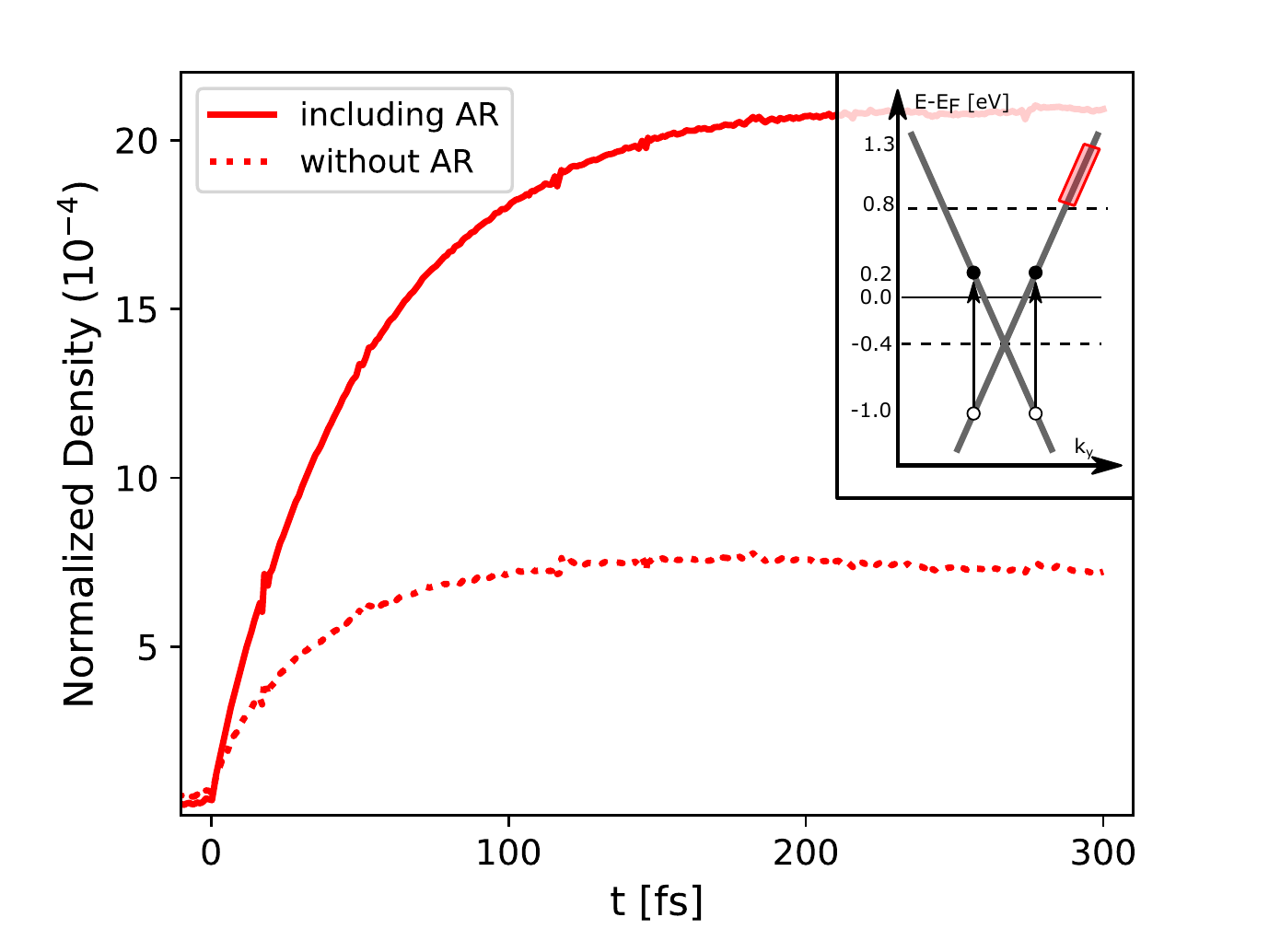}
    \caption{Computed time-dependent carrier densities integrated over the energy range $E-E_F>0.8$~eV corresponding to the experimentally accessible population of high-energy electrons. The density is normalized to the total density of electrons in the conduction band. The red solid line is computed including all scattering channels and the dashed line excluding Auger processes.}
    \label{fig:theo_plot}
\end{figure}


In order to obtain theoretical support of the experimental fingerprints of AR processes, we carried out simulations of two-body Coulomb scattering processes in graphene at the level of Boltzmann scattering integrals (see SI for details). Exemplary results of these simulations are shown in Fig.~\ref{fig:theo_plot}. To keep the simulation and comparison simple, we initiated the dynamics in the simulation ($\Delta t=0$~fs) by anisotropically redistributing the charge carriers between the conduction and the valence band. The resulting imbalance of electrons and holes at $E-E_F=0.2$~eV and $-1.0$~eV corresponds to the experimental excitation, as shown in the inset of Fig.~\ref{fig:theo_plot}. The main part of Fig.~\ref{fig:theo_plot} shows transient electron populations for  $E-E_F>0.8$~eV calculated for two cases: (i) including all Coulomb-mediated e-e scattering pathways, and (ii) for an artificial case neglecting all scattering events in which one carrier crosses bands. Thus in scenario (ii), AR processes are excluded in the simulation. The simulation shows that the total carrier density at $E-E_F>0.8$~eV, i.e., at energies beyond those reached by the pump pulse rises after the excitation to a value that is only a small fraction of the total carrier density in both cases. We do not observe a subsequent decrease of the density in this energy range because we do not include electron-phonon interactions that would lead to such a ``cooling'' behavior. However, the direct comparison clearly shows that AR processes play a large enough role at high energies such that its signatures can be picked up by the experiment. While our timescales are slightly longer than those in the experiment due to our choice of the screening parameter, we conclude from the calculations that it is the large influence of AR processes on the high-energy dynamics that makes them detectable, even though the total carrier density at high-energies is extremely small.




In conclusion, we have reported on the first direct experimental observation of Auger recombination in graphene. 
We show that strong optical excitation of charge carriers in n\nobreakdash-doped graphene will lead to significant AR, resulting in a macroscopic number of highly excited charge carriers at energies higher than reached by the optical excitation itself. Depending on the dominant scattering processes, a distinct temporal structure on a sub-50-fs timescale has been identified and can be used in the future for further studies on primary thermalization events in graphene. For example, it will be highly interesting to systematically vary the doping level in order to control the relative contribution of AR, IE or other e-e scattering processes, and thereby tune the response to the optical excitation.

\section{ACKNOWLEDGEMENTS}

This work was funded by the Deutsche Forschungsgemeinschaft (DFG, German Research Foundation) - 217133147/SFB 1073, projects B07 and B03. G.S.M.J. and M.R. acknowledge funding by the Alexander von Humboldt Foundation. S.S. acknowledges the Dorothea Schl\"ozer Postdoctoral Program for Women. D.M.P acknowledges support from the Joint Research Project "GIQS" (18SIB07). This project has received funding from the EMPIR programme co-financed by the Participating States and from the European Union’s Horizon 2020 research and innovation program.

\section{REFERENCES}

\end{document}


\title{
Supplemental Information:\\
Direct Access to Auger Scattering in Graphene
}

\author{Marius~Keunecke} \email{mkeunec@gwdg.de}%
\address{I. Physikalisches Institut, Georg-August-Universit\"at G\"ottingen, Friedrich-Hund-Platz 1, 37077 G\"ottingen, Germany}

\author{David~Schmitt} %
\address{I. Physikalisches Institut, Georg-August-Universit\"at G\"ottingen, Friedrich-Hund-Platz 1, 37077 G\"ottingen, Germany}

\author{Marcel~Reutzel} \email{marcel.reutzel@phys.uni-goettingen.de} %
\address{I. Physikalisches Institut, Georg-August-Universit\"at G\"ottingen, Friedrich-Hund-Platz 1, 37077 G\"ottingen, Germany}

\author{Marius~Weber} %
\address{TU Kaiserslautern, Erwin-Schrödinger-Str. 46, 67663 Kaiserslautern, Germany}

\author{Christina~Möller}%
\address{I. Physikalisches Institut, Georg-August-Universit\"at G\"ottingen, Friedrich-Hund-Platz 1, 37077 G\"ottingen, Germany}

\author{G.~S.~Matthijs~Jansen} %
\address{I. Physikalisches Institut, Georg-August-Universit\"at G\"ottingen, Friedrich-Hund-Platz 1, 37077 G\"ottingen, Germany}

\author{Tridev~A.~Mishra}%
\affiliation{Institut für Theoretische Physik, Georg-August-Universit\"at G\"ottingen, Friedrich-Hund-Platz 1, 37077 G\"ottingen, Germany}

\author{Alexander~Osterkorn}%
\affiliation{Institut für Theoretische Physik, Georg-August-Universit\"at G\"ottingen, Friedrich-Hund-Platz 1, 37077 G\"ottingen, Germany}

\author{Wiebke~Bennecke}%
\affiliation{I. Physikalisches Institut, Georg-August-Universit\"at G\"ottingen, Friedrich-Hund-Platz 1, 37077 G\"ottingen, Germany}

\author{Klaus~Pierz} %
\address{Physikalisch-Technische Bundesanstalt, Bundesallee 100, 38116 Braunschweig, Germany}

\author{Hans~Werner~Schumacher} %
\address{Physikalisch-Technische Bundesanstalt, Bundesallee 100, 38116 Braunschweig, Germany}

\author{Davood~Momeni~Pakdehi} %
\address{Physikalisch-Technische Bundesanstalt, Bundesallee 100, 38116 Braunschweig, Germany}


\author{Daniel~Steil} %
\address{I. Physikalisches Institut, Georg-August-Universit\"at G\"ottingen, Friedrich-Hund-Platz 1, 37077 G\"ottingen, Germany}

\author{Salvatore~R.~Manmana}%
\affiliation{Institut für Theoretische Physik, Georg-August-Universit\"at G\"ottingen, Friedrich-Hund-Platz 1, 37077 G\"ottingen, Germany}

\author{Sabine~Steil} 
\address{I. Physikalisches Institut, Georg-August-Universit\"at G\"ottingen, Friedrich-Hund-Platz 1, 37077 G\"ottingen, Germany}

\author{Stefan~Kehrein}%
\affiliation{Institut für Theoretische Physik, Georg-August-Universit\"at G\"ottingen, Friedrich-Hund-Platz 1, 37077 G\"ottingen, Germany}

\author{Hans~Christian~Schneider} %
\address{TU Kaiserslautern, Erwin-Schrödinger-Str. 46, 67663 Kaiserslautern, Germany}

\author{Stefan~Mathias} \email{smathias@uni-goettingen.de}%
\address{I. Physikalisches Institut, Georg-August-Universit\"at G\"ottingen, Friedrich-Hund-Platz 1, 37077 G\"ottingen, Germany}

\maketitle

The supplemental material contains additional TR-MM data excluding contributions of two-photon transitions, and sideband yield to the identification of AR. In addition, we provide details on the simulations.

\section{Excluding contributions of two-photon excitation processes}

\begin{figure}[hbt!]
    \centering
    \includegraphics[width=\textwidth]{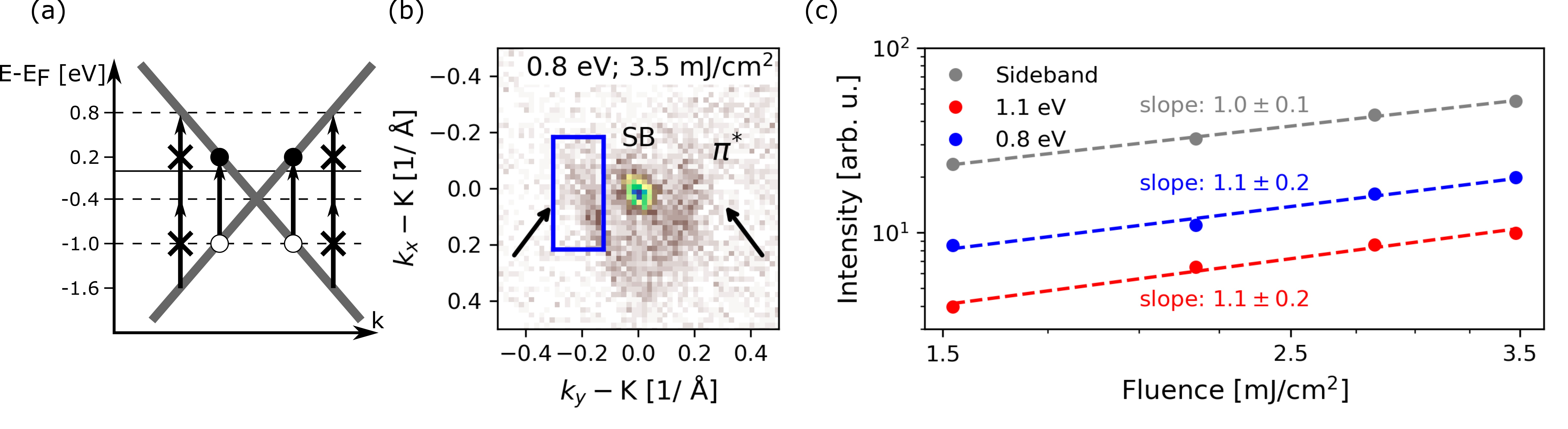}
    \caption{Fluence dependence of the charge carrier occupation in the conduction band obtained with $p$-polarized pump light.
    (a) Excitation diagram for possible two-photon processes (crossed arrows) that would terminate at $E-E_F=0.8$~eV.
    (b) Momentum map at $E-E_F=0.8$~eV showing the anisotropic population of the conduction band at $\Delta t=0$~fs, as well as the contribution of the sideband (SB) centered at the $K$-point. The colored boxes indicate the regions of interest evaluated in (b).
    (c) Fluence dependence of the photoemission yield in the sideband (black), at $E-E_F=0.8$~eV where two-photon transitions would terminate, and at $E-E_F\geq0.8$~eV where we identify direct evidence for AR. The corresponding intensities have
been corrected for background counts.}  
    \label{fig:fluence}
\end{figure}

In order to exclude the possible contribution of two-photon excitation processes to the occupation of high-energy regions of the conduction band [$E-E_F\geq0.8$~eV; excitation diagram in Fig.~\ref{fig:fluence}~(a)], we performed fluence-dependent TR-MM experiments [Fig.~\ref{fig:fluence}~(b,c)]. 
We make use of the sideband photoemission intensity, which is expected to scale linearly with pump fluence~\cite{saathoff_laser-assisted_2008}. As expected, the sideband intensity scales with a slope of $1.0\pm0.1$ on the log-log scale, confirming the linear dependence.  
In particular at $E-E_F=0.8$~eV, the energy for resonant two-photon excitation, the photoemission yield does not scale quadratically with the fluence, but with an exponent of $1.1 \pm 0.2$.
We can therefore conclude that two-photon absorption does not contribute to the signal at 0.8~eV, and that the majority of the detected charge carriers must have been excited through e-e scattering processes.
Furthermore, when evaluating energies above 0.8~eV up to 1.4~eV, i.e. the energy window that provides direct evidence for AR, we also extract a slope of $1.1 \pm 0.2$. 

\section{Excluding the influence of sideband intensities on carrier scattering analysis ($p$-polarized pump light)}

From Fig.~3~(a) of the main text, one can see that the sideband trace crosses the main Dirac cone where optical excitation proceeds resonantly (black ROI). We calibrate $\Delta t = 0$~fs in the energy- and momentum-region where the sideband does not overlap with the main Dirac cone (grey ROI). In Fig.~3~(b,c), the grey ROI and the black ROI have distinctly different $\Delta t_{\rm max}$. Such a scenario is not be expected if significant contributions of the black ROI would stem from sideband yield; instead, the dynamics measured in the black ROI are dominantly determined by the charge carrier dynamics in the conduction band. If still considering minor contributions of the sideband to the black ROI, $\Delta t_{\rm max}=8\pm1$~fs has to be treated as the lower limit. Note, in addition, that in the next section we unambiguously exclude the contribution of the sideband yield to the observed dynamics based on measurements with $s$-polarized pump light.


\section{Additional data for s-polarized pump pulses}
\begin{figure}[hbt!]
    \centering
    \includegraphics[width=.5\textwidth]{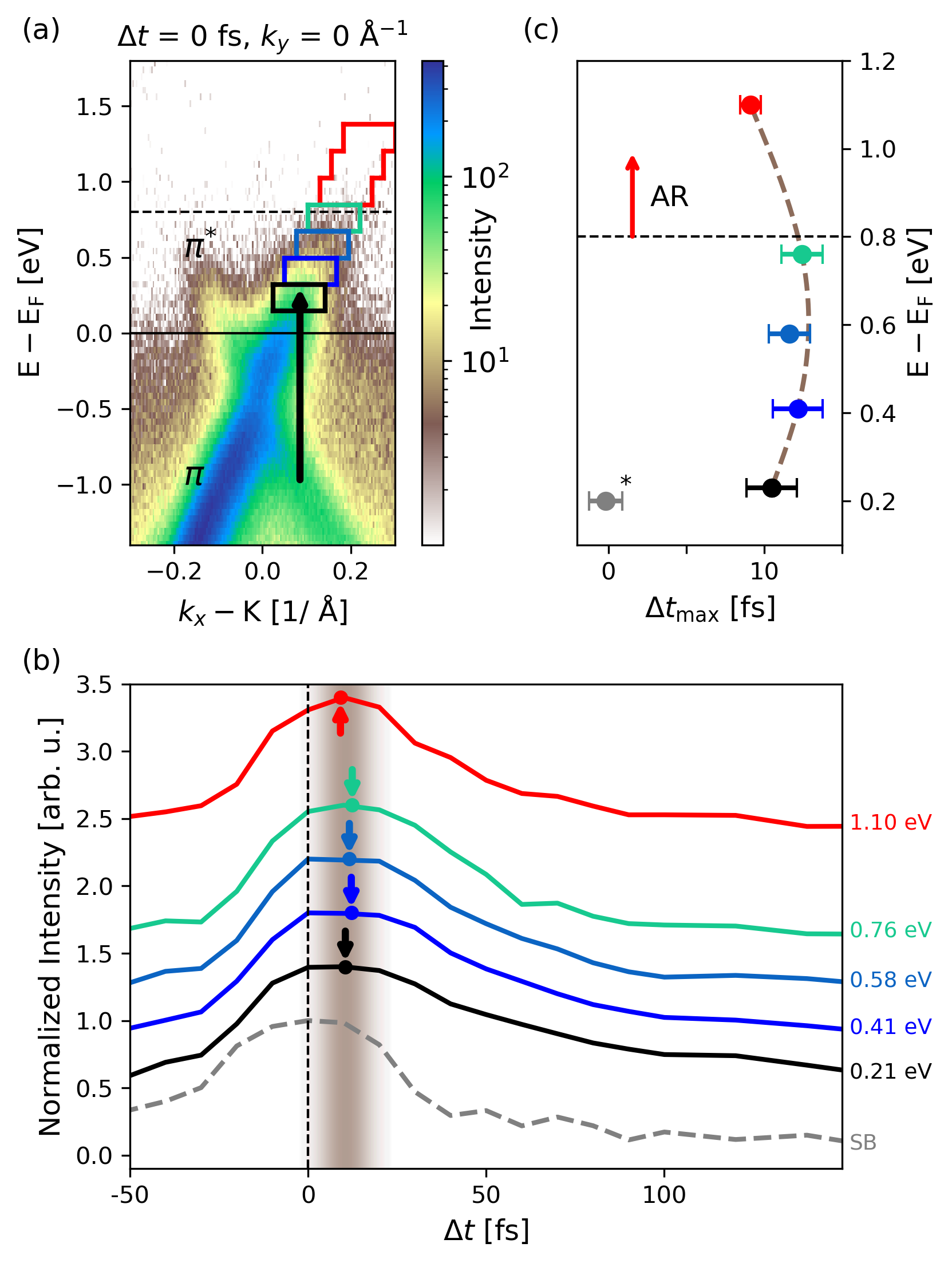}
    \caption{Identification of AR in the ultrafast, energy- and momentum-resolved, redistribution of non-thermal charge carriers in the conduction band of graphene. The same data processing as in Fig.~3 of the main text is carried out, but for $s$-polarized pump light.
    (a) ($E$, $k_y$)-cut at $k_x=0$. The regions of interest evaluated in (b) and (c) are indicated by colored boxes. The dotted line at $E-E_F=0.8$~eV indicates the threshold energy that can be reached with a single inter- or intraband scattering event. Higher energies can only be reached by a single AR, or multiple inter- and intraband scattering events.
    (b) The photoemission yield in the ROIs is evaluated as a function of pump-probe delay. The time of maximum charge carrier occupation ($\Delta t_{\rm max}$) in the respective ROI is indicated with a colored arrow. Note that the SB labelled time trace is obtained from a weak sideband signal. 
    (c) Energy-dependence of $\Delta t_{\rm max}$ obtained by Gaussian fitting of the time traces between -20~fs and +30~fs; error bars correspond to the 1$\sigma$ standard deviation of the fit. The sideband trace (grey) is used for the calibration of the time-axis (note that its plotting on the $E-E_F$-axis is not physical and therefore marked with a $\**$). For increasing $E-E_F$, $\Delta t_{\rm max}$ first increases, saturates, and finally decreases again for energies where AR is expected to become the dominant scattering event. The brown dashed line is a guide to the eye through the data.}  
    \label{fig:s-pol}
\end{figure}

Figure~\ref{fig:s-pol} shows TR-MM data obtained on n-doped graphene for $s$-polarized pump light (1.2~eV, 17.3~mJ/c$\mathrm{m}^2$). The data is evaluated as done in the main text for the $p$-polarized case in Fig.~3. Similar as discussed in the main text, we observe a distinct shift of $\Delta t_{\rm max}$ = $9\pm1$ with energy and momentum that we interpret as a direct evidence for AR. 

Note that when probing the Dirac cone located in the optical plane, for $s$-polarized pumping, no contributions from the laser-assisted photoelectric effect~\cite{saathoff_laser-assisted_2008} can be expected, as detailed in Ref.~\cite{Keunecke20prb}. The fact that the $s$-polarized data and analysis in Fig.~\ref{fig:s-pol} shows the same energy-dependent delay further supports that the intensity of the sideband in the p-polarized case does not contribute significantly to our analysis procedure.
\clearpage

\section{Simulation of Scattering Dynamics}

\subsection{Boltzmann Scattering}
We outline the general treatment of electron-electron scattering for a solid in the single particle picture.
We calculate the time evolution of the electronic distribution functions $f^{\nu}_{\boldsymbol{k}}$ in state $|\nu,\boldsymbol{k}\rangle$, where $\nu$ is the band index, by using the Boltzmann scattering integral~\cite{Bonitz_1996}
\begin{equation}
\begin{split}
\frac{d}{dt}f_{\boldsymbol{k}}(t) =&\frac{2L^{2n}}{\hbar^2(2\pi)^n}\int d^n{q}\int d^{n}l \; \big[V_{\mu_2,\mu_3}^{\nu,\mu_1}(\boldsymbol{q})\big]^{2}\Upsilon(\Delta E,t,\Gamma)\\ 
&[1-f^{\nu}_{\boldsymbol{k}}][1-f^{\mu_1}_{\boldsymbol{l}+\boldsymbol{q}}]f^{\mu_2}_{\boldsymbol{l}}f^{\mu_3}_{\boldsymbol{k}+\boldsymbol{q}}-f^{\nu}_{\boldsymbol{k}}f^{\mu_1}_{\boldsymbol{l}+\boldsymbol{q}}[1-f^{\mu_2}_{\boldsymbol{l}}][1-f^{\mu_3}_{\boldsymbol{k}+\boldsymbol{q}}] 
\end{split}
\label{collision-2}
\end{equation}
which describes the two-fermion scattering transition $\boldsymbol{l} \rightarrow \boldsymbol{l}+\boldsymbol{q}$ and $\boldsymbol{k}+\boldsymbol{q} \rightarrow \boldsymbol{k}$ mediated by the screened Coulomb interaction.
The energy difference $\Delta E$ between initial and final states enters here via
\begin{equation}
\Upsilon(\Delta E, t, \Gamma)=\frac{\hbar}{\Gamma^2+\Delta E^2} \Big(\Big[\Delta E \sin\Big(\frac{\Delta E}{\hbar} (t-t_0)\Big)-\Gamma\cos\Big(\frac{\Delta E}{\hbar}(t-t_0) \Big)\Big]e^{-\frac{\Gamma}{\hbar}(t-t_0)}+\Gamma \Big).
\label{verbreiterte_Delta_funktion}
\end{equation}
where $\Gamma$ is the  imaginary part of the self-energy. This expression approaches $\Upsilon \to \hbar \delta(\Delta E)$ at longer times and for $\Gamma \to 0$, which conserves kinetic energy, but at short times it leads to an increase of the kinetic energy of our system during the building up of correlations.~\cite{morawetz} In order to have a computationally feasible model, the self energy will be assumed as constant $\Gamma$ . A finite constant $\Gamma$ also leads to an increase in energy and we choose $\Gamma= 0.001 \mathrm{meV}$ keep this effect small for the duration of our simulations. Equation \eqref{collision-2} can be derived from a self-consistent quantum kinetic Boltzmann equation by integrating over retardation effects. \cite{Bonitz_1996,Haug}

\subsection{Graphene}

For the numerical solution of the dynamical equation for the distribution we employ some simplifications, which are described below.
The bandstructure is  linearized around the $\Gamma$ point\cite{Knorr_Book} with a slope of 650\,meV/nm, and the matrix elements are calculated using the linearized dispersions. The Coulomb matrix elements can then be written as
\begin{equation} 
V_{\mu_2,\mu_3}^{\nu,\mu_1}=V_q^{2D}g_{\mu_2,\mu_3}^{\nu,\mu_1}
\label{eq:V-graphene}
\end{equation}
and we use a statically screened Coulomb potential $V_q^{2D}$. We have assumed that an additional momentum dependent factor in Eq.~\eqref{eq:V-graphene} can be dropped,~\cite{Knorr_Book} and we use a background dielectric constant of 4 and a constant screening parameter $\kappa = 4\text{nm}^{-1}$. 
The coefficient $g_{\mu_2,\mu_3}^{\nu,\mu_1}$ is determined by
\begin{align}
g_{\mu_2,\mu_3}^{\nu,\mu_1}=\frac{1}{4}\left(1+c_{\nu,\mu_2}\frac{e(\boldsymbol{k}_1)^{*}e(\boldsymbol{k}_3)}{|e(\boldsymbol{k}_1)e(\boldsymbol{k}_3)|}\right)\left(1+c_{\mu_1,\mu_3}\frac{e(\boldsymbol{k}_2)^{*}e(\boldsymbol{k}_4)}{|e(\boldsymbol{k}_2)e(\boldsymbol{k}_4)|}\right).
\end{align}
The prefactor $c_{a,b}$ equals 1 for $a=b$ (intraband processes) and $-1$ for $a\neq b$ (interband processes), further
$e(\boldsymbol{k})=-\frac{a_{0}\sqrt{3}}{2}(ik^x+k^y) $.
These matrix elements describe all scattering channels near the Dirac cone. To quantify the impact of Auger processes we are interested in index combinations $\mu_1=\mu_2=\mu_3 \neq \nu$ and all their permutations. Within these eight possible combinations, Auger recombination and impact excitation are included, both types of Auger scattering having basically the same strength.

We have made two further important simplifications: In order to focus on electron electron scattering, we did not include electron-phonon scattering  processes, which would result in a decrease of the kinetic energy. The excitation is included in a purely phenomenological way by instantaneously creating the excited carrier density.

\subsection{Numerical Details}

Finally we provide a short review of the used numerical scheme and the boundary conditions.
We implemented Eq.~\eqref{collision-2} for a Cartesian $k$ grid, which has the advantage that it does not introduce numerical errors in the carrier density conservation. Therefore we are able to calculate the scattering properties of the complete 2D k-space with an arbitrary excitation. Due to the numerical costly right-hand side of  Eq.\eqref{collision-2} we used parallelization techniques and a Dormand-Prince differential equation solver with adaptive step size. Finally the k-space was sampled with 51 grid points in each direction and with a maximal momentum of $k_{\text{max}}=2.5 \mathrm{nm^{-1}}$.
This approach using a Cartesian $k$ grid conserves carrier density by construction, which is important as we are looking 
relatively small effects at high energies (> 800 meV). On the other hand, it is numerically quite costly.
The initial distribution is chosen as a Fermi-Dirac distribution $f_{\pm}^{\text{eq}}=1/(1+e^{[(E_{\pm}-\mu) /(k_{\text{B}}T)]}) $
with a chemical potential of 400 meV to model the relaxed electrons introduced by doping.

We assume an instantaneous excitation of the form
\begin{equation}
   \delta f= A e^{-\frac{1}{b}(\boldsymbol{k}-\boldsymbol{k}_e)^2}|\sin(k_x)| \label{df}
\end{equation}
with the amplitude $A$, the width of the excitation $b$ and the center of the excitation $\boldsymbol{k}_e$, which follows the spacial shape of the measured excitation and the calculations from Ref.~\cite{Winzer_2013}. In order to distinguish between scattering effects and building up of correlations, we let the system with the doped carrier density relax and then initialize the calculation by changing the electronic distributions by $\pm \delta f$ in the upper/lower band with positive/negative energy dispersion.

\begin{figure}
    \centering
    \includegraphics[scale=0.8]{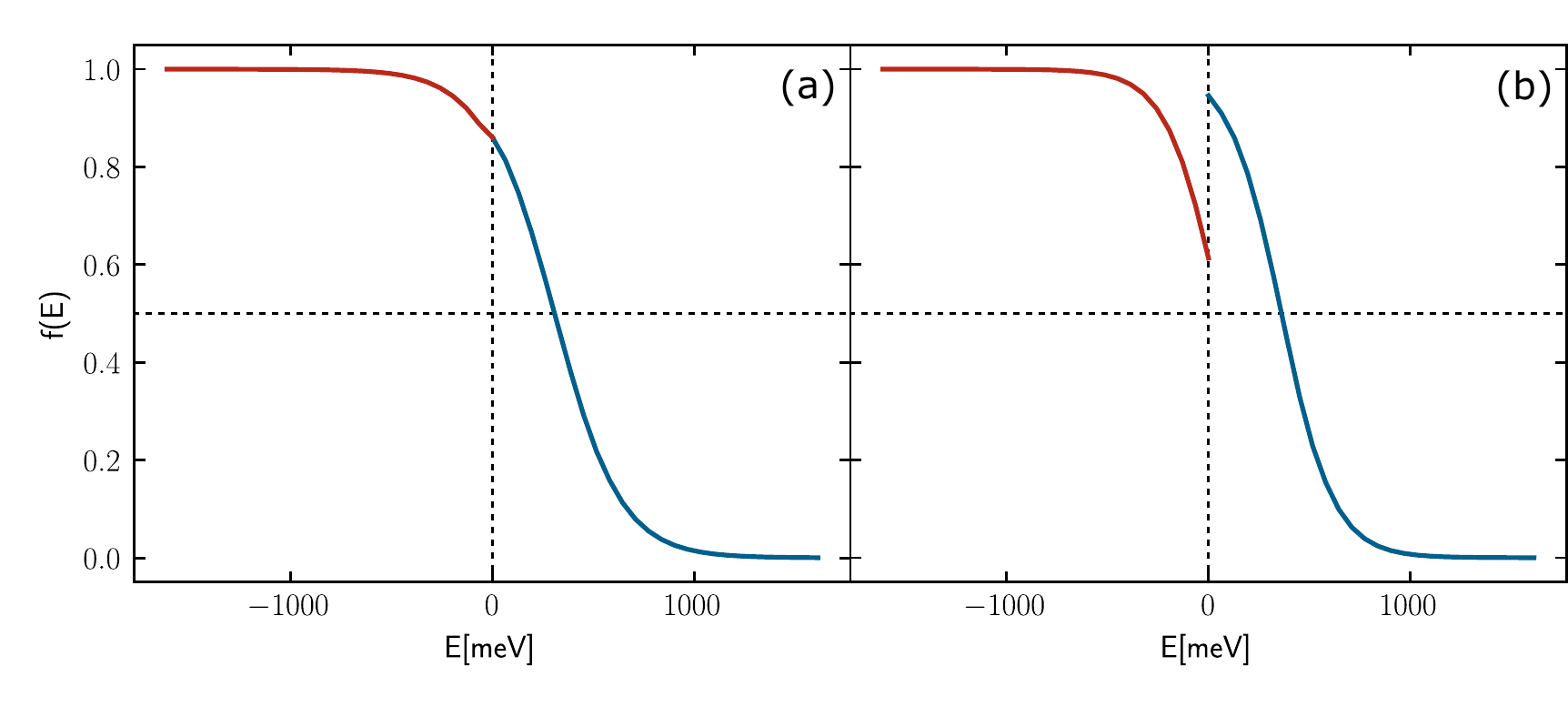}
    \caption{Distribution functions after 300 fs, for the full calculation (a), and the calculation without Auger processes. The red/blue part corresponds to the lower/upper band. The full calculation reaches a quasi-equilibrium with a chemical potential of about 300\, meV. Neglecting Auger processes in the calculation prevents the system from reaching a quasi equilibrium.}
    \label{fig-supplement-distributions}
\end{figure}

To eliminate Auger type processes, we exclude the Coulomb matrix elements relevant for these transitions by setting $g_{\mu,\mu}^{\mu,\nu}=0$. We can thus compare the scattering dynamics with and without Auger processes.
In order to replicate the photoemission experiments to some extent, we integrate the time-dependent distributions over the  $(0,k_y)$ axis for the energy interval $0.8$-$1.3$ eV to obtain the time dependent density traces shown in Fig. 4 in the main text. A numerical test case of the dynamics is presented in Fig.~\ref{fig-supplement-distributions}, where we compare the energy-dependent distributions obtained 300\,fs after the instantaneous excitation. If all scattering processes are included the system reaches a Fermi-Dirac distribution with a higher temperature and a different chemical potential, as the excitation increases the kinetic energy of carriers, and electron-electron scattering does not dissipate this energy. Switching the Auger scattering contributions off leads to the distributions in Fig.~\ref{fig-supplement-distributions}(b), which are clearly not equilibrated.

\clearpage